\documentclass[twocolumn,showpacs,floats,prb,aps,showpacs]{revtex4}
\usepackage[final]{graphicx}
\usepackage[dvipsnames]{color}
\usepackage{dcolumn}
\usepackage{hhline}
\DeclareGraphicsExtensions{.pdf,.eps}

\begin{document}

\title{Charge-transfer excitations in molecular donor-acceptor complexes within the many-body Bethe-Salpeter approach}

\author{X. Blase and C. Attaccalite}

\affiliation{ Institut N\'{e}el, CNRS and Universit\'{e} Joseph Fourier,
B.P. 166, 38042 Grenoble Cedex 09, France.}

\date{\today}

\begin{abstract}
We study within the perturbative many-body  $GW$ and Bethe-Salpeter approach the low 
lying singlet charge-transfer excitations in molecular donor-acceptor complexes associating 
benzene, naphtalene and anthracene derivatives with the tetracyanoethylene acceptor. 
Our calculations demonstrate that such techniques can reproduce the experimental 
data with a mean average error of 0.1-0.15 eV for the present set of dimers, in excellent 
agreement with the best time-dependent density functional studies with optimized 
range-separated functionals.  The present results pave the way to the study of photoinduced 
charge transfer processes in photovoltaic devices with a parameter-free \textit{ab initio} 
approach showing equivalent accuracy for finite and extended systems.
\end{abstract}
                                  

\maketitle


A large set of studies aiming at clarifying the processes assisting the separation of the photogenerated 
electron-hole pairs in  organic  solar cells revealed the importance of charge transfer excitations at 
donor/acceptor interfaces.  \cite{CTCrefs}  From a theoretical point of view, the  fundamental issue of
correctly modelling, at the quantum mechanical level, the electron-hole interaction,
is a long standing difficulty. In particular, while standard density functional theory \cite{tddft} 
can accurately describe excitations with a strong spatial overlap between 
the involved occupied and unoccupied states, the interaction of weakly overlapping electron-hole pairs requires 
the development of novel functionals with a fine tuning of the local and non-local exchange and correlation 
contributions, \cite{longrange} a strategy followed e.g. in the recent developments around the range-separated 
functionals. \cite{Stein09} 


In this work, we study within the many-body $GW$ and Bethe-Salpeter ($BSE$) perturbation framework 
the charge-transfer excitation energies of ten donor/acceptor molecular complexes, associating tetracyanoethylene (TCNE) 
with benzene, naphtalene, anthracene and their derivatives.  We show that these techniques provide results which are in 
excellent agreement with available experimental data. The accuracy is equivalent to that of the best time-dependent 
density functional theory based on optimized range-separated functionals. We emphasize in particular the importance 
of self-consistency when performing $GW$-$BSE$ calculations starting from DFT calculations with (semi)local
functionals.

\begin{figure}
\begin{center}
\includegraphics*[width=0.45\textwidth]{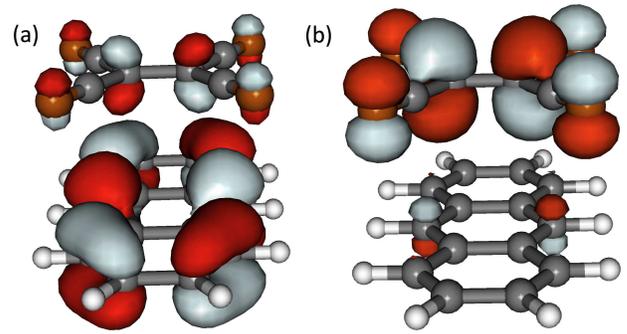}
\caption{ (Color online) Isocontour representation of  the (a) HOMO and (b) LUMO states of the
anthracene-TCNE complex. The contours have been set to 20$\%$ of the maximum
value of the wavefunction. Blue and red contours indicate different signs of the 
wavefunction. The grey, white and brown atoms are carbon, hydrogen and nitrogen, respectively.
 The $\pi$/$\pi^*$ character of the wavefunctions is apparent.}
\label{fig1}
\end{center}
\end{figure}

The choice of such donor-acceptor complexes combining TCNE and acene derivatives is dictated by the availability
of gas phase experiments in the case of benzene, toluene, o-xylene and naphtalene donors, \cite{Hanazaki72}
complemented by experiments in solution for anthracene and its derivatives. \cite{Masnovi84}
Such experiments provide invaluable reference data for the study of the merits of the various theoretical
approaches, even though experiments performed in solution complicate the comparison with theoretical calculations 
performed on the isolated complexes. 

We adopt the geometries obtained by Stein and coworkers at the DFT-B3LYP level in their TDDFT study of the 
same systems using an optimized range-separated functional labeled BNL($\gamma$=0.3) in the following. 
\cite{Stein09} To illustrate the typical geometry of such systems, and highlight their donor-acceptor 
character, we plot in Fig.~1 an isocontour representation of the highest occupied  (HOMO) and lowest 
unoccupied (LUMO) molecular orbital of the anthracene-TCNE complex, showing the clear localization 
of the HOMO (LUMO) on the anthracene donor (TCNE acceptor).

As compared to the TDDFT approach, the $GW$-$BSE$ formalism proceeds in two steps. First, the 
quasiparticle HOMO-LUMO gap, namely the difference between the ionization energy and electronic 
affinity, is calculated within the $GW$ approximation. 
\cite{Hedin65,Strinati80,Hybertsen86,Godby88,Onida02,Aulbur}
Traditionally, such calculations are
performed non-self-consistently, namely as a ``single-shot" correction to the starting DFT-LDA 
Kohn-Sham eigenstates, an approach labeled $G_0W_0$ in the following. While this approach 
has been shown since the mid-eighties to yield excellent results for most extended semiconducting 
systems, \cite{Aulbur} several studies \cite{Hahn05,Rostgaard10,Blase11,Faber11}
evidenced that in the case of small molecular systems, the $G_0W_0$(LDA)
approximation yields too small ionization potential and HOMO-LUMO gaps.  
It was shown that such a problem could be cured either by starting from Hartree-Fock eigenstates, 
\cite{Hahn05,Rostgaard10,Blase11} or by performing a simple self-consistent update of the 
quasiparticle energies when building the self-energy operator, \cite{Blase11,Faber11} an approach 
labeled $GW$ in what follows. 

In a  second step, the neutral (optical) excitation energies can be obtained as the eigenvalues of
the Bethe-Salpeter $H^{e-h}$ Hamiltonian \cite{Sham66,Strinati82,Rohlfing98,Shirley98,Albrecht98} 
which, in the  $|\phi_{i}^e \phi_j^h>$ product basis of the 
non-interacting unoccupied $|\phi_{i}^e>$ and occupied  $|\phi_j^h>$  single-particle states, is
composed of three terms as follows:

\begin{eqnarray*}
  H^{diag}_{ij,kl} &=& \delta_{i,k} \delta_{j,l} \left(  \varepsilon_i^{QP} - \varepsilon_j^{QP} \right) \\ 
  H^{direct}_{ij,kl} &=& - <\phi_i^e \phi_j^h| W(r,r') |\phi_{k}^e \phi_l^h> \\
  H^{exchange}_{ij,kl} &=& 2 <\phi_i^e \phi_j^h| V^C(r,r') |\phi_{k}^h \phi_l^e>
\end{eqnarray*}

\noindent with ($\varepsilon_i^{QP}$,$\phi_i^{e/h}$) the best single-particle eigenstates in the absence of
electron-hole interaction.  $W(r,r')$ and $V^C(r,r')$ are the (statically) screened and bare Coulomb potential 
respectively.  Following the above discussion, the $\varepsilon_i^{QP}$ can be taken to be the $G_0W_0$(LDA) 
or $GW$ quasiparticle energies. Clearly, a too small HOMO-LUMO gap in $H^{diag}$ will result in too small 
neutral excitation energies. 


Our $G_0W_0$-$BSE$ and $GW$-$BSE$ calculations are performed with the {\sc{Fiesta}} code \cite{Blase11} 
which exploits an even-tempered auxiliary basis representation of the two-point operators such as $V^C $ and $W$, while 
the input DFT-LDA eigenstates are taken to be those of the {\sc{Siesta}} DFT code \cite{siesta} using a large
 triple-zeta plus double polarization (TZ2P) basis.  In the case of the smaller complexes with benzene, toluene 
and o-xylene, these calculations are double-checked using the planewave basis set\cite{planewaves}.


\begin{figure}
\begin{center}
\includegraphics*[width=0.45\textwidth]{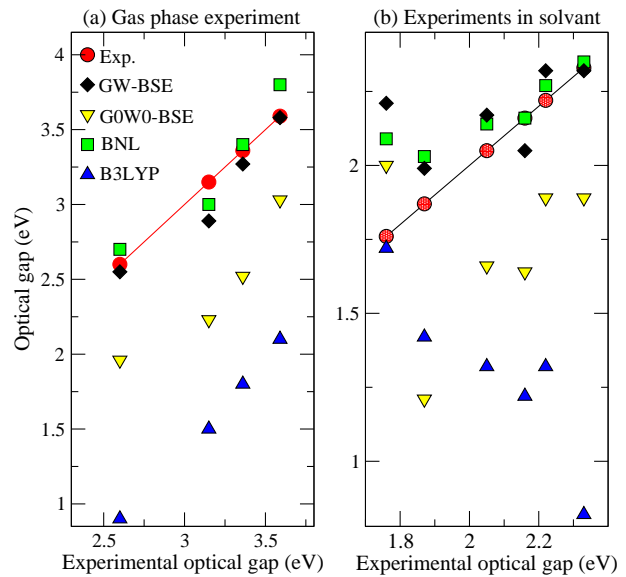}
\caption{
Experimental and theoretical first singlet excitation energies in eV.  The signification of the
symbols is given in (a). Experimental data are from (a) Ref.~\onlinecite{Hanazaki72} for gas phase
experiments on benzene, toluene, o-xylene and naphtalene donors, and (b) Ref.~\onlinecite{Masnovi84} 
for experiments in solution on anthracene and its derivatives. Following Stein and coworkers in 
Ref.~\onlinecite{Stein09},  a constant 0.32 eV shift has been added to the experimental data in (b) 
to (tentatively) mimic the bathochromic shift. The B3LYP and BNL data are the theoretical TDDFT 
results from Ref.~\onlinecite{Stein09}. Figures (a) and (b) are not on the same energy scale.} 
\label{fig2}
\end{center}
\end{figure}

We compile in Fig.~2 and Table I our calculated excitation energies within the $G_0W_0$(LDA)-BSE (yellow down 
triangle) and $GW$-BSE (black diamonds) approaches, that we compare to the experimental data (red circles) 
and to the TDDFT results of Stein and coworkers \cite{Stein09} with the B3LYP (blue up triangle) and range-separated 
BNL($\gamma$=0.3) (green squares) functionals.  Following the analysis by Stein and coworkers, \cite{Stein09} experimental
data obtained in solution (light red circles) have been shifted upwards by a 0.32 eV constant that approximately 
accounts for the solvent induced bathochromic shift. As shown in the Table, our results as obtained with the 
{\sc Fiesta} code and with the planewave Yambo approach (in parenthesis),  are in good agreement within 0.1 eV 
accuracy, showing that the present results are independent of the technicalities of the various implementations.

\begin{table}
\begin{tabular}{c|c|c|c|c}
\hline
\hline
   donor   & $G_0W_0$-$BSE$ &   $GW$-BSE   &  BNL & Exp.  \\
\hline
benzene    &  3.03 (3.10) &    3.58  (3.63)   &   3.8                 &  3.59   \\
toluene    &  2.52 (2.67) &    3.27  (3.37)   &   3.4                 &  3.36   \\
o-xylene   &  2.23 (2.28) &    2.89  (3.00)   &   3.0                 &  3.15   \\
naphtalene &  1.96        &    2.55     &         2.7                 &  2.60   \\
\hline
MAE        &   0.74  (0.68)   &    0.10 (0.07)     &   0.12                &   - \\
\hline
\hline
 substituent & \multicolumn{4}{c}{anthracene derivatives}  \\
\hline
none      &   1.66        &    2.17    &   2.14          &  (2.05)  \\
9-carbo-methoxy &  1.64   &    2.05    &   2.16          &  (2.16)  \\
9-cyano   &   1.89        &    2.32    &   2.35          &  (2.33)  \\
9-methyl    &      1.21   &    1.99    &    2.03         &    (1.87)   \\
9,10-dimethyl  &   2.00   &    2.21    &    2.09         &   (1.76)    \\
9-formyl       &   1.89   &    2.32    &    2.27         &   (2.22)    \\
\hline
MAE            &   0.43   &    0.15    &   0.11          &   - \\
\hline
\hline
\end{tabular}
\caption{Experimental and theoretical optical gap (eV) for  donor-TCNE complexes.
$G_0W_0$-$BSE$ and $GW$-BSE results in parenthesis have been obtained with a 
planewave basis implementation (see text). MAE is the mean absolute error. 
The BNL column reports the TDDFT results with optimized range-separated BNL($\gamma$=0.3) 
functionals from Ref.~\onlinecite{Stein09}.  Experimental results are from 
Ref.~\onlinecite{Hanazaki72,Masnovi84}. The experimental results in parenthesis 
were obtained in solution and a  0.32 eV constant energy has been added (see text).}
\label{table}
\end{table}

The mean absolute error (MAE) as compared to experiments is found within the $GW$-$BSE$ approach to 
be 0.10 eV for ``gas phase"  benzene, toluene, o-xylene and naphtalene complexes, and 0.13 eV averaging over all 
systems, including the corrected data from the experiments in solution.  An interesting and satisfactory 
result is the good agreement with the previous TDDFT calculations based on range-separated functionals, with 
a localization parameter $\gamma$=0.3 (see Ref.~\onlinecite{Stein09}) optimized to reproduce the correct 
quasiparticle band gap.  In the two cases where theory and experiment start showing a severe disagreement, 
namely with methyl and dimethyl anthracene derivatives (the two lowest experimental excitation energy in Fig.~2b), 
both TDDFT-BNL and $GW$-$BSE$ results agree in overestimating the extrapolated experimental data.  This may indicate 
that the estimation of the solvation effect is incorrectly accounted for by the rigid 0.32 eV shift.  Finally, 
as emphasized above, an important observation is that the $G_0W_0$-$BSE$ approach provides significantly 
too small excitation energy, as a result of too small a $G_0W_0$-LDA quasiparticle band gap.\cite{tricklastra} 
As expected, the  $GW$-$BSE$ results are much better than those obtained with the TDDFT-B3LYP approach.\cite{b3lypweird}

In conclusion, we have shown that $GW$-$BSE$ first-principles many-body perturbation theory provides charge
transfer excitation energies in excellent agreement with experiment, with a mean average error of the order of 
0.1-0.2 eV.  The present results are of equivalent accuracy as compared to the best range-separated functionals
TDDFT calculations, with an approach showing equivalent merits for finite size and extended systems without
any adjustable parameter. 
In fact the screened Coulomb potential $W(r,r')$ through which the holes and electrons are interacting 
is a non-local operator with a decay in finite  size or extended systems automatically adjusted 
through the evaluation of the dielectric properties of the system.
The analysis of such interaction in donor-acceptor complexes may help in the
future to better understand the features of the best exchange-correlation kernels within the TDDFT framework.

\textbf{Acknowledgements.} Computing time has been provided by the local CIMENT and national GENGI-IDRIS supercomputing centers in Grenoble and Orsay (project No.~100063), respectively. 


\begin{thebibliography}{150bliography}
\bibitem{planewaves}
GW calculations in planewaves have been performed  by using the Yambo package,\cite{Yambo} taking as
an input the DFT-LDA Kohn-Sham states provided by the Quantum-Espresso package.\cite{pwscf} 
We have employed a 17.5~$\AA$ cubic supercell, a truncated Coulomb interaction in order to avoid spurious
interaction between the repeated images and a plasmon pole model fitted to the dielectric constant at imaginary
frequencies 0 and 1 Hartree.
\bibitem{b3lypweird}
Even though TD-B3LYP results for anthracene and its derivatives may seem anticorrelated with experiment, 
there are better correlated with theoretical values, except noticeably for the 9-cyano case where
TD-B3LYP dramatically underestimates the transition energy (see Ref.~\onlinecite{Stein09}).
\end{thebibliography}
\end{document}